\documentclass{article}

\usepackage{PRIMEarxiv}

\usepackage[authoryear, round]{natbib}
\usepackage{amsmath}
\usepackage{amsthm}
\newtheorem{assumption}{Assumption}

\usepackage[utf8]{inputenc} 
\usepackage[T1]{fontenc}    
\usepackage{hyperref}       
\usepackage{url}            
\usepackage{booktabs}       
\usepackage{amsfonts}       
\usepackage{nicefrac}       
\usepackage{microtype}      
\usepackage{lipsum}
\usepackage{fancyhdr}       
\usepackage{graphicx}       
\graphicspath{{media/}}     

\title{Networks And Productivity – A Study In Economic Scholars During COVID-19
}

\author{
  Hanqiao Zhang \\
  University of California Irvine \\
  Irvine, CA\\
  \texttt{hanqiaoz@uci.edu} \\
   \And
  Joy D. Xiuyao Yang \\
  Florida Institute of Technology \\
  Melbourne, FL\\
  \texttt{xyang@fit.edu} \\
}

\begin{document}
\maketitle

\begin{abstract}
The COVID-19 pandemic has disrupted traditional academic collaboration patterns, offering a unique opportunity to analyze the influence of peer effects and collaboration dynamics on research productivity. Using a novel network dataset, this paper investigates the role of peer effects on the productivity of economists, as measured by their publication count, during both pre-pandemic and pandemic periods. The results indicate that peer effects were significant in the pre-pandemic period but not during the pandemic. Additionally, the study sheds light on gender and race differences. These findings enhance our understanding of how research collaboration impacts knowledge production and provide insights that may inform policies aimed at promoting collaboration and boosting research productivity within the academic community.
\end{abstract}


\section{Introduction}

Academic productivity is a critical metric in decisions related to academic hiring, promotions, and evaluations of research programs and departments \citep{gingras2016bibliometrics, bornmann2008citation}. However, measuring academic productivity is a complex question. In general, a higher number of publications is often indicative of a scholar’s active research involvement and significant contributions to the field of study. Recently, there has been an increasing focus on identifying factors that enhance the productivity of scholars across various disciplines. Among these, the role of peer effects and co-authorship has been recognized as significantly influential \citep{ductor2014social}. Scholars with robust social networks and interactions within their field are likely to publish more due to greater access to resources, including research funding, data, and other materials. Co-authorship further promotes knowledge sharing, potentially sparking new research ideas and increasing publication opportunities. Notably, \citep{petersen2015quantifying} highlights how a single strong connection can significantly boost a scholar’s productivity and citation rates.

The COVID-19 pandemic has fundamentally altered the way we live, work, and collaborate. As it forced the closure of universities and research institutions, the pandemic inadvertently reshaped the landscape of academic research collaboration. This unprecedented situation offers a unique opportunity to examine the peer effects and co-authorship dynamics among economic scholars during both pre-pandemic and pandemic periods. Working from home has potentially transformed scholars’ approaches to writing, communication, and collaboration. On one hand, the flexibility of schedules and greater autonomy over their work have allowed for enhanced virtual collaboration through video conferencing and instant messaging among scholars located in different regions or time zones. On the other hand, working from home may lead to greater isolation and fewer opportunities for informal interactions and discussions with co-authors and potential co-authors. It may also introduce distractions and disruptions, such as the challenge of balancing work responsibilities with home duties.

This paper utilizes a novel dataset derived from Google Scholar, encompassing scholarly literature and co-authorship networks of economic scholars. We estimate peer effects on the number of publications by these scholars during both pre-pandemic and pandemic periods. The study holds significant implications for several reasons. Firstly, it sheds light on the role of co-authorship in academic productivity, offering insights into how research collaborations contribute to knowledge production and dissemination. Secondly, by contrasting the periods before and during the COVID-19 crisis, this study enhances our understanding of how the pandemic has altered academic collaboration patterns and overall scholarly productivity in economics. Lastly, the findings also illuminate gender and race differences, offering insights that may guide policies aimed at enhancing collaboration and increasing research productivity within the academic community.

This paper is organized as follows: Section 2 reviews the relevant literature on peer effects and academic productivity. Section 3 describes the data and provides summary statistics. Section 4 outlines the model framework and the empirical strategy employed. Section 5 discusses the main results. Finally, Section 6 concludes the paper and suggests directions for future research.

\section{Literature Reivew}

From a methodological perspective, our study aligns closely with two main branches of the literature. Firstly, it contributes to the rapidly expanding body of research on peer effects within networks. As illustrated by \citep{10.2307/2298123}, peer influences manifest in two distinct forms: endogenous peer effects, which pertain to the influence of peers’ outcomes, and contextual peer effects, which relate to the influence of peers' characteristics. Distinguishing between these effects is challenging due to the simultaneity of behaviors among interacting agents. Pioneering studies on network interactions such as \citep{bramoulle2009identification,de2010identification,lin2010identifying,laschever2005doughboys} lay the groundwork in this field. For instance, \citep{bramoulle2009identification} establishes the benchmark linear-in-means model of peer effects, detailing the identification conditions under the assumption that peers' influence does not extend beyond direct connections. This assumption—that agents’ characteristics influence individual outcomes solely through their effect on peers’ outcomes—provides valid instruments for addressing correlated effects, thus allowing for the identification of both endogenous and contextual peer effects. A critical insight is that the identification of these effects heavily depends on the network structure. Subsequent research by \citep{bramoulle2013comment, arduini2014identification, arduini2020treatment, beugnot2019gender} explores heterogeneous peer effects, noting differential impacts among genders. \citep{masten2018random} extends this analysis by incorporating heterogeneity, proposing that coefficients for endogenous peer effects vary randomly within a linear-in-means model, and demonstrating that these variations can be precisely identified in the absence of contextual effects for any exogenous characteristic.

The second strand of literature focuses on addressing the issue of correlated effects and leveraging the identification possibilities offered by interaction networks. Researchers have developed at least four broad strategies to tackle this issue: random peers, random shocks, structural endogeneity, and panel data. The first strategy, random peers, involves peers who are randomly allocated through natural or designed experiments. For example, \citep{sacerdote2001peer} examines roommate pairings at Dartmouth College, \citep{falk2006clean} pairs workers randomly in a laboratory setting, and \citep{de2010identification} investigates major selections among undergraduates at Bocconi. The key insight here is that with random peers, an agent’s observed and unobserved characteristics are uncorrelated with those of their peers. The second strategy, random shocks, is exemplified by studies like \citep{dieye2014accounting}, which use exogenous variations for treatment randomization within a linear-in-means framework to identify causal impacts despite potential endogeneity in the network. Other studies, such as \citep{miguel2004worms, kremer2007illusion, crepon2013labor}, explore spillover effects, highlighting the complexities of estimating causal impacts when outcomes may depend on a comprehensive array of treatments. The third strategy, termed the structural framework, was initially proposed by \citep{goldsmith2013social} to manage correlated effects by potentially controlling for network endogeneity in peer effect regressions. This approach, reminiscent of Heckman’s correction, aims to account for sample selection and unobserved common factors simultaneously. The fourth strategy pertains to the use of panel data to analyze peer effects, although this aspect of the literature is relatively undeveloped. A few studies, such as \citep{patnam2011corporate, comola2021treatment, de2020consumption}, have employed panel data incorporating individual fixed effects. However, they generally do not tackle contextual peer effects related to time-invariant characteristics, marking a significant limitation and an important area for future research.

The model employed in the present study belongs to the structural endogeneity framework. It is related to models that use static games with incomplete information, in which agents act non-cooperatively (see \citep{harsanyi1967games, osborne1994course}). The assumption of incomplete information in peer effect models for discrete outcomes is widely studied (e.g., \citep{brock2001discrete, bajari2010estimating, yang2017social, de2017econometrics}). In the literature, an agent $i$’s decision is influenced by their own observable characteristics, unobservable individual type, and other agents’ choices. Recent studies by \citep{boucher2020estimating, houndetoungan2022count} propose methods to estimate the network’s probability distribution using cross-sectional data when the network is imperfectly observed. They construct a network game where each agent chooses an integer outcome to maximize their preference, which includes observed characteristics of the agent and peers, the difference between the agent's choice and that of peers, a cost function, and a private signal. They prove that under certain assumptions, there is a unique Bayesian Nash Equilibrium for this game, and they propose an estimator based on pseudo-likelihood maximization.

The literature of empirical papers exploring the relationship between co-authorship and academic productivity has expanded recently. Nonetheless, consensus remains elusive regarding whether this relationship is positive, negative, or insignificant. For example, \citep{cainelli2015strength} demonstrates that economists who are more collaborative are also more productive, and factors such as tenure, age, and geographical variables do not have a significant impact on productivity. Similarly, \citep{ductor2015does} finds a positive correlation between intellectual collaboration and individual performance after accounting for endogenous network formation, unobservable heterogeneity, and factors that vary over time. As indirect evidence, \citep{bosquet2013large} identifies that, at the individual level, the average publication quality rises with the average number of authors per paper, individual field diversity, the total number of published papers, and the presence of foreign co-authors, while female and older academics tend to publish less frequently.

Conversely, some scholars argue that the relationship between co-authorship and productivity exhibits a negative correlation or lacks statistical significance. \citep{hollis2001co} discovers that for a specific scholar, increased co-authorship leads to higher quality, longer, and more frequent publications. However, after adjusting for the number of authors, the relationship between co-authorship and a scholar's attributable output becomes negative. Other research contends that after controlling for article length, journal and author quality, subject area, and scholar fixed effects, the productivity of prior collaborators is not a significant determinant of a researcher’s own productivity \citep{cheng2022productivity} or higher quality research \citep{medoff2003collaboration}. Furthermore, \citep{oettl2012reconceptualizing} finds that co-authors of highly helpful scientists who die experience a decrease in output quality but not output quantity.

\section{Economic Scholars Data}

The Economic Scholars dataset consists of 1,671 core faculty members from the top 50 Economics Schools in the United States, based on the 2022 US News rankings, who have registered a homepage on Google Scholar. The dataset excludes visiting professors, teaching professors, and lecturers. These homepages provide rich information on the individuals' influences and research journeys in academia. Besides the names and affiliations, scholars may list their research interests and sub-fields at the top of the page. On the right-hand side, the scholar's cumulative citations, H-index, and I10-index are displayed, along with regular coauthors and a histogram showing the number of citations and publications each year for the last 10 years. On the left-hand side, a comprehensive list of the scholar's academic output can be viewed, including the paper's title, journal, authors, year, and the number of citations. See a snapshot of the Google Scholar homepage in the figure below.

\begin{figure}[!htbp]
  \caption{Google Scholar Homepage Example}
  \centering
  \includegraphics[width=0.8\textwidth]{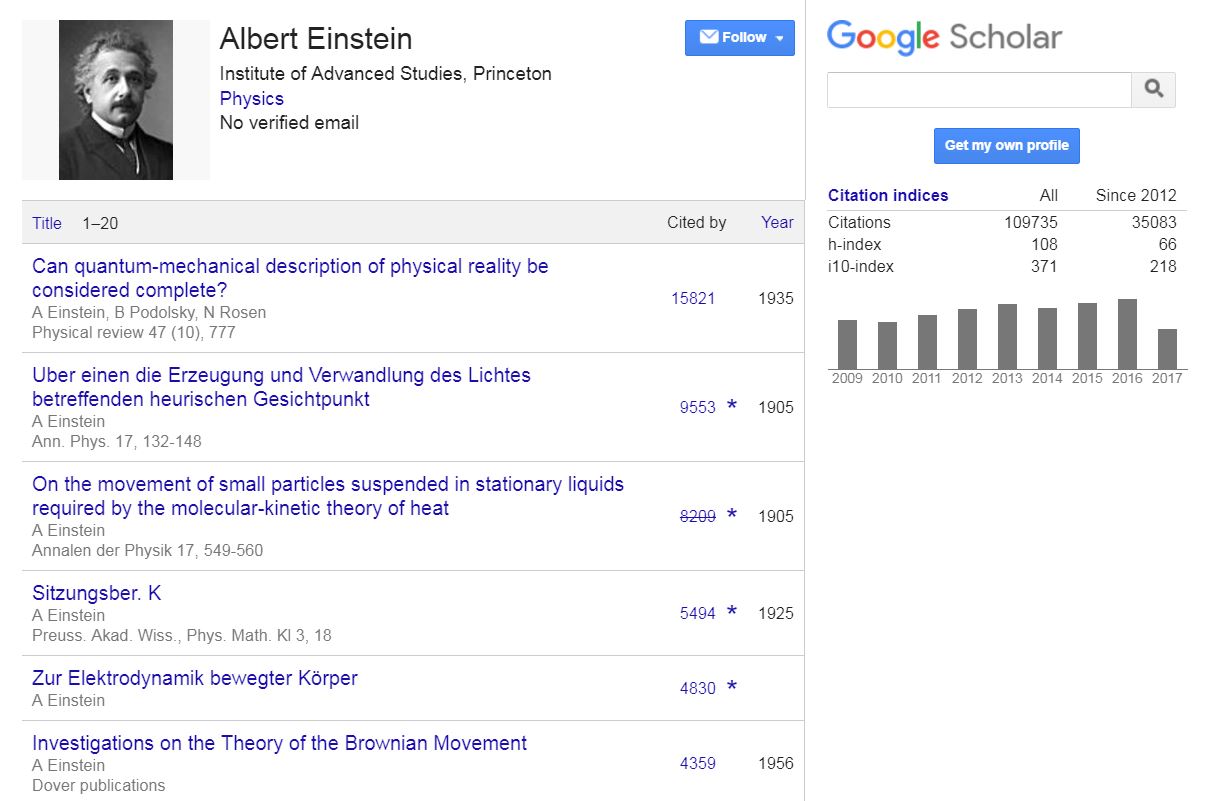}
\end{figure}

A key feature of the data collected from Google Scholar is its comprehensiveness. The website aggregates information from a wide range of sources, including journal articles, conference papers, theses and dissertations, technical reports, books and book chapters, patents, working papers, and repositories. Thus, compared to using data sources from academic journals, the lagged effect of co-authorship on scholars' academic output due to journals' reviewing processes can be largely alleviated.

Based on scholars' publications, we construct an economics co-authorship network for each year over the last 10 years. Each node represents a faculty member in the economics department, and there is an edge between any two nodes if they have co-authored at least one paper in a given year. The edge density of a temporal network indicates the prevalence of scholar collaborations in that year, or the percentage of observed co-authorships over all possible collaborations between any two scholars. Economic scholars' productivity, as revealed by the annual average number of publications, gradually increased from 2013, surged to a peak in 2020, and then slowly declined in 2021 and 2022. Network edge density reached its highest level in 2018, declined to its lowest point in 2020, and then rebounded in 2021 and 2022. Interestingly, scholars' average productivity moved in the same direction as the prevalence of collaborations before 2018 but in the opposite direction from 2019 to 2021.

\begin{figure}[!htbp]
  \caption{Average Number of Publications and Edge Density by Year}
  \centering
  \includegraphics[width=0.5\textwidth]{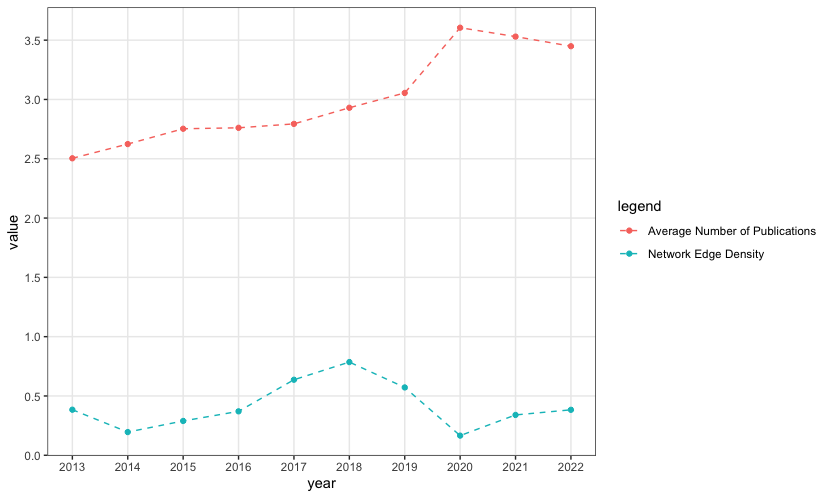}
\end{figure}

The experience of economic faculty in academia averages 25 years as of 2023, measured by the difference between 2022 and the year of the scholar's first published article on Google Scholar. African American scholars, identified either directly by their listed nationality in CVs or indirectly by the combination of the predicted country of origin using names and faculty pictures, account for 0.9\% of the economics faculty. In our sample, 18\% of the faculty are female.

We also label the expertise of economic scholars with their listed sub-fields on their Google Scholar homepage. For those who have left this section blank, we fill in the missing values using the fields shown on the department faculty website and the scholar's curriculum vitae. The sub-fields are not mutually exclusive, and it is reasonable for scholars to have multiple specialties. For example, a macroeconomist may specify econometrics, monetary policy, international economics, or finance as their expertise. In our sample, economic/econometric theory, macroeconomics, and labor economics are the three top fields that scholars work on, followed by econometrics, industrial organization, and development economics.

\begin{table}[!htbp] \centering 
  \caption{Summary Statistics of Economic Scholars} 
\begin{tabular}{@{\extracolsep{5pt}}lccccc} 
\\[-1.8ex]\hline 
\hline \\[-1.8ex] 
Statistic & \multicolumn{1}{c}{Mean} & \multicolumn{1}{c}{St. Dev.} & \multicolumn{1}{c}{Min} & \multicolumn{1}{c}{Median} & \multicolumn{1}{c}{Max} \\ 
\hline \\[-1.8ex] 
Years in Academia & 24.96 & 15.97 & 1 & 22 & 72 \\ 
African American & 0.009 & 0.09 & 0 & 0 & 1 \\
Female & 0.18 & 0.39 & 0 & 0 & 1 \\
Number of Publications in 2013 & 2.504 & 3.154 & 0 & 2 & 25 \\ 
Number of Publications in 2014 & 2.624 & 3.195 & 0 & 2 & 30 \\ 
Number of Publications in 2015 & 2.75 & 3.38 & 0 & 2 & 27 \\ 
Number of Publications in 2016 & 2.76 & 3.31 & 0 & 2 & 21 \\ 
Number of Publications in 2017 & 2.79 & 3.45 & 0 & 2 & 40 \\ 
Number of Publications in 2018 & 2.93 & 3.43 & 0 & 2 & 40 \\ 
Number of Publications in 2019 & 3.06 & 3.84 & 0 & 2 & 33 \\ 
Number of Publications in 2020 & 3.61 & 4.93 & 0 & 2 & 68 \\ 
Number of Publications in 2021 & 3.53 & 4.46 & 0 & 2 & 55 \\ 
\hline \\[-1.8ex] 
\end{tabular} 
\end{table} 

\begin{table}[!htbp] \centering 
  \caption{Sub-fields of Economic Scholars}
  \setlength\tabcolsep{3pt}
\begin{tabular}{lc}
\\[-1.8ex]\hline \hline \\[-1.8ex] 
\\[-1.8ex] & \multicolumn{1}{c}{Economics Sub-field (\%)} \\ 
\hline \\ [-1.8ex]
  Theory & 18.25 \\ 
  Macroeconomics & 19.80 \\
  Labor Economics & 18.01 \\
  Econometrics & 15.92 \\
  Industrial Organization & 11.31 \\
  Development Economics & 11.61 \\
  Health Economics & 7.60 \\
  Financial Economics & 10.29 \\
 \hline \\[-1.8ex] 
\hline \\[-1.8ex]
\end{tabular} 
\end{table}

\section{Network Game with Peer Effects and Incomplete Information}

Following the notation in \citep{houndetoungan2022count}, suppose there is a population of $n$ agents that interact through a network matrix $\mathbf{G}$ with zero diagonal and non-negative elements $g_{ij}$ that represents the proximity of $i$ and $j$. Each agent $i$ chooses an integer outcome, $y_i$, to maximize his or her individual utility:

\begin{equation}
    \nonumber
    U_i\left(y_i,\mathbf{y}_{-i}\right) = \psi_i y_i - c\left(y_i\right) - \frac{\lambda}{2}\left(y_i-\bar{y_i}\right)^2 + e_i\left(y_i\right)
\end{equation}

In the equation, $\mathbf{y}_{-i}=\left(y_1,\cdots,y_{i-1},y_{i+1},\cdots,y_n\right)$. Observable characteristics of $i$ and his or her peers is contained in $\psi_i=\mathbf{x}_i\mathbf{'\beta}+\mathbf{\bar{x_i}'\gamma}$, where $\mathbf{x}_i$ is the vector of the agent's observable characteristics and $\mathbf{\bar{x_i}}$ is the average characteristics of the peers. Own effects $\mathbf{\beta}$ and contextual effects $\mathbf{\gamma}$ are parameters to be estimated. Function $c\left(\cdot\right)$ captures the cost of choosing $y_i$, and $\bar{y_i}=\sum_{j=1}^n g_{ij}y_j$. The peer effect, $\lambda \geq 0$, is designed to show conformity so that $\frac{\lambda}{2}\left(y_i-\bar{y_i}\right)^2$ represents the social cost that is greater if the difference between the choice of agent $i$ and his or her peers is larger. $\left(e_i\left(r\right)\right)_{r\in\mathbb{N}}$ is a random variable sequence that indicates the agent's private type. The value is observable to $i$ for any $r \in \mathbb{N}$, but not to other agents. Since the types and thus the choices of other agents are not observable, agent $i$ would maximize the expectation of his or her preferences conditional on the information set $I_i=\left\{ \psi_i, \mathbf{\psi}_{-i}, \mathbf{g}_i, \mathbf{G}_{-i} \right\}$:

\begin{equation}
    \nonumber
    U_i^e\left(y_i\right) = \psi_i y_i - c\left(y_i\right) - \frac{\lambda}{2} 
\mathbb{E}_{\bar{y_i}|I_i} \left[\left(y_i-\bar{y_i}\right)^2\right] + e_i\left(y_i\right)
\end{equation}

Define $\Delta$ as the first difference operator. It is proved in \citep{houndetoungan2022count} that under the following three assumptions, there is a unique integer $r_0$ that maximizes the preference $U_i^e\left(\cdot\right)$, and $U_i^e\left(r\right) \geq \max\left\{ 
U_i^e\left(r-1\right),U_i^e\left(r+1\right) \right\}$ if and only if $r=r_0$:

\begin{assumption}
  $c(\cdot)$ is a strictly convex and increasing function.
\end{assumption}
\begin{assumption}
  For any $r \in \mathbb{N}$, $e_i\left(r\right)=e_i\left(r-1\right)+\epsilon_i$, where $\epsilon_i|I_i$ are independent and identically follow a continuous symmetric distribution with cdf function $F_{\epsilon|I}$, and pdf function $f_{\epsilon|I}$.
\end{assumption}
\begin{assumption}
  $\lim\limits_{r \to \infty} r^{-\rho} \left( \Delta c(r+1) - \Delta c(r) \right)>0$, and $f_{\epsilon|I}(x)=o(|x|^{-\kappa})$ at $\infty$, where $\rho \geq 0$, $(1+\rho)(\kappa-1)>2$.
\end{assumption}

The first assumption implies that $\Delta c(r+1)-\Delta c(r)>0$. The expected payoff is  strictly concave and has a global maximum that could be reached at a single point. The second assumption suggests that agents consider the same information $\epsilon_i$ for any additional $r$ so that $\Delta e_i(r)$ does not depend on $y_i$. The third assumption suggests that when $y_i$ is sufficiently high, the cost increases at a minimum rate. The tail of $f_{\epsilon|I}(x)$ needs to decay, and the trade-off condition between $\rho$ and $\kappa$ guarantees that when $r\rightarrow\infty$, the probability of $y_i=r$ converges to $0$ at some rate.

Agent $i$ chooses $r$ if and only if $U_i^e\left(r\right) \geq U_i^e\left(r-1\right)$ and $U_i^e\left(r\right) \geq U_i^e\left(r+1\right)$. Substituting $U_i^e\left(\cdot\right)$ and $e_i\left(\cdot\right)$ into the two conditions, we have $-\psi_i-\lambda\bar{y_i}^e+a_r \leq \epsilon_i \leq -\psi_i-\lambda\bar{y_i}^e+a_{r+1}$, where $a_r=\Delta c(r)+\lambda r-\frac{\lambda}{2}$, $\bar{y_i}^e=\sum_{j=1}^n g_{ij}y_j^e$, $y_i^e$ is agent $i$'s rational expected choice conditional on information set $I_i$. The probability of agent $i$ choosing $r$, $p_{ir}$, could readily be written as:

\begin{equation}
    \nonumber
    p_{ir} = F_{\epsilon|I}\left( \lambda\bar{y_i}^e + \psi_i - a_r \right) - F_{\epsilon|I}\left( \lambda\bar{y_i}^e + \psi_i - a_{r+1} \right)
\end{equation}

The expected outcome associate with the belief system $\mathbf{p}=(p_{ir})$ could be written as $y_i^e=\sum_{r=1}^\infty rp_{ir}=\sum_{r=1}^\infty F_{\epsilon|I}\left( \lambda\bar{y_i}^e + \psi_i - a_r \right)$. Although the expected payoff has a global maximum, it is possible that there are multiple expected outcomes and belief systems $\mathbf{p}$. To avoid the multiple rational expected equilibria issue, a threshold for the peer effect needs to be imposed:

\begin{assumption}
  $\lambda < B_c/\vert\vert\mathbf{G}\vert\vert_{-\infty}$, where $B_c= \left( \max_{u\in\mathbb{R}} \sum_{r=1}^\infty f_{\epsilon|I}\left(u-a_r\right) \right)^{-1}$
\end{assumption}

With the above four assumptions, this game is proved to have a unique Bayesian Nash Equilibrium given by $\mathbf{y^*}=(y_1^*,\cdots,y_n^*)'$, where $y_i^*$ is the maximizer of the expected payoff $U_i^e\left(\cdot\right)$. 

In the payoff function $\psi_i=\mathbf{x}_i\mathbf{'\beta}+\mathbf{\bar{x_i}'\gamma}$, let the observable characteristics of agent $i$ and the peers, $\mathbf{x}_i$ and $\mathbf{\bar{x_i}}$, be $1 \times K$ vectors. Specify $\mathbf{X}=\left[\mathbf{x}_1,\cdots,\mathbf{x}_n\right]'$ as a $n \times K$ matrix, and $\mathbf{\psi=Z\Gamma}$ where $\mathbf{Z}=\left[\mathbf{X\ GX}\right]$ and $\mathbf{\Gamma}=\left(\mathbf{\beta}',\mathbf{\gamma}'\right)'$. Define $\delta_r=a_r-a_{r-1}$ for $r \geq 2$ and $\delta_1=0$. As $a_r=\Delta c(r)+\lambda r-\frac{\lambda}{2}$, $\delta_r=\Delta\Delta c(r)+\lambda$ for $r \geq 2$. Since $c(\cdot)$ is non-parametric, an infinite number of $\delta_r$ needs to be estimated. For identification purposes, the paper assumes the limitation in \textbf{Assumption 3} is reached for large $r$:

\begin{assumption}
  There exists a positive constant $R$, such that $\forall r > R$, $\delta_r=(r-1)^\rho\bar{\delta}+\lambda$, where $\bar{\delta}>0$, $\rho>0$.
\end{assumption}

The probability of agent $i$ choosing $r$ could then be re-written as:
\begin{equation}
    \nonumber
    p_{ir} = F_{\epsilon|I}\left( \lambda\bar{y_i}^e + \mathbf{z}_i'\mathbf{\Gamma} - a_r \right) - F_{\epsilon|I}\left( \lambda\bar{y_i}^e + \mathbf{z}_i'\mathbf{\Gamma} - a_{r+1} \right)
\end{equation}
where $\mathbf{z}_i'$ is $\mathbf{Z}$'s $i$-th row, $a_0=-\infty$, $a_r=a_1+\sum_{k=1}^r\delta_k$ for all $r \geq 1$, $\delta_1=0$, $\delta_r=(r-1)^\rho \bar{\delta}+\lambda$ for all $r>R$. Let $\bar{R}$ be the smallest $R$ for which \textbf{Assumption 5} holds. It is shown that under a few additional assumptions, if $F_\epsilon$ is known, then $\lambda,\mathbf{\Gamma},\mathbf{\delta},\bar{\delta},\rho$ are point identified.

Parameter estimation proceeds with a likelihood approach. Assume $\epsilon_i$ follows a standard normal distribution, then the probability $p_{ir}$ would be:
\begin{equation}
    \nonumber
    p_{ir} = \Phi\left( \lambda \mathbf{g}_i\mathbf{y}^e + \mathbf{z}_i'\mathbf{\Gamma} - a_r \right) - \Phi\left( \lambda \mathbf{g}_i\mathbf{y}^e + \mathbf{z}_i'\mathbf{\Gamma} - a_{r+1} \right)
\end{equation}
where $\Phi(\cdot)$ is the CDF of standard normal distribution, $\mathbf{L}$ is a mapping, $\mathbf{y}^e=\mathbf{L}(\mathbf{\theta,y}^e)=(l_1(\mathbf{\theta},\mathbf{y}^e),\cdots,l_n(\mathbf{\theta},\mathbf{y}^e))'$, and $l_i(\mathbf{\theta},\mathbf{y}^e)=\sum_{r=1}^\infty \Phi\left( \lambda \mathbf{g}_i\mathbf{y}^e + \mathbf{z}_i'\mathbf{\Gamma} - a_r \right)$. For any fixed $\bar{R}$, since $\mathbf{y}^e$ is not observed,
it needs to be computed for every value of $\mathbf{\theta}$. Alternatively, the parameters could be estimated using the nested pseudo likelihood (NPL) algorithm in \citep{aguirregabiria2007sequential}, which takes advantage of an iterative process. The algorithm maximizes a pseudo-likelihood function:
\begin{equation}
    \nonumber
    L_n\left(\mathbf{\theta,y}^e\right)=\frac{1}{n} \sum_{i=1}^n\sum_{r=0}^\infty d_{ir}log(p_{ir})
\end{equation}
where $\mathbf{\theta}=\left( log(\lambda), \mathbf{\Gamma}', log(\Tilde{\mathbf{\delta}}'),log(\bar{\delta}), log(\rho) \right)'$, $d_{ir}=1$ if $y_i=r$ and $0$ otherwise. It starts by guessing a set of initial probabilities for each agent's choices, and then updating these probabilities in each iteration until the parameters and probabilities converge to a stable solution, e.g. $||\mathbf{\theta}_{(t)}-\mathbf{\theta}_{(t-1)}||_1$ and $||\mathbf{y}_{(t)}-\mathbf{y}_{(t-1)}||_1$ are less than $10^{-4}$.

The model also takes into account the endogeneity problem induced by agents' unobserved characteristics. For instance, in our application, scholars' familiarity with coding, or their extent to communicate with others could affect both which scholars they may collaborate with ($\mathbf{G}$) and their own number of publications ($\mathbf{y}$). Let the latent utility of scholar $i$ and $j$ being coauthors be $\mathbf{g}_{ij}^*=\mathbf{\ddot{x}}_{ij}\mathbf{\bar{\beta}}+\mu_i+\nu_j+\eta_{ij}$, where $\mathbf{\ddot{x}}_{ij}$ contains dyadic variables, $mu_i$ and $\nu_j$ are individual fixed effects. The probability of scholar $i$ and $j$ being coauthors could be modelled as:

\begin{equation}
    \nonumber
    P_{ij}= \mathbb{P}(\mathbf{g}_{ij}^*>0) = F_\eta \left( \mathbf{\ddot{x}}_{ij}\mathbf{\bar{\beta}}+\mu_i+\nu_j \right)
\end{equation}

The fixed effects are assumed to be unobservable for the researcher, but observable for the scholars so that they are included in the information set.

\begin{assumption}
    For continuous function $h_\epsilon$, $\epsilon_i=h_\epsilon(\mu_i,\nu_i,\bar{\mu}_i,\bar{\nu}_i)+\epsilon_i^*$, where $\epsilon_i^*$ is independent of $\mathbf{Z}$ and $\mathbf{G}$, $\bar{\mu}_i=\sum_{j=1}^n\mathbf{g}_{ij}\mu_j$, $\bar{\nu}_i=\sum_{j=1}^n\mathbf{g}_{ij}\nu_j$.
\end{assumption}

The $\epsilon$ could be replaced by $h_\epsilon(\mu_i,\nu_i,\bar{\mu}_i,\bar{\nu}_i)+\epsilon_i^*$. Adapting \textbf{Assumption 2} and \textbf{Assumption 3} to $\epsilon_i^*$, the defined Bayesian Nash Equilibrium is still valid. The estimator computes $\hat{\mu}_i$ and $\hat{\nu}_i$ using a standard Logit in the first step, then substitute the estimated values for $\mu_i$ and $nu_i$ in the second step. Function $h_\epsilon$ is approximated using a sieve method. We refer readers that are interested in the model and technical details to the original paper \citep{houndetoungan2022count}.

The Economics scholar data contains the annual number of publications for each scholar from 2018 to 2021. We split the sample into pre-Covid and Covid periods (2018-2019 and 2020-2021) and apply the model to each of these 2-year periods to estimate the peer effect. Specifically, We define $y_i$ as the total number of publications of scholar $i$ within each 2-year period. Data is segmented in this way because productivity and collaboration are observed to move in different directions before and after 2019. The pandemic and schools' transition to virtual learning in 2020-2021 also inevitably affected scholars' productivity and collaboration, so the peer effect $\lambda$ is expected to have some change. We keep the length of each period the same to ensure that the estimated results in the two models are comparable.

We include gender, an indicator for African American scholars, expertise, recent productivity indicated by the average number of publications of the scholar in the previous 3 years, total citations up to the first year of each period, and academic experience in the observable characteristics. We discretize scholars' recent productivity, total citations, and academic experience, setting scholars with an average of 0 or 1 publication in the previous 3 years, fewer than 100 citations up to the first year of each period, and less than 10 years of experience as the reference levels respectively.

In the presence of a productivity rise during Covid times, we construct an additional Covid index for the period 2019-2021 to control for the extent to which scholars' publications after the outbreak of the epidemic are related to the Covid context. We use topic modeling to predict the topic of each paper published within these three years based on the title and abstract. The model is capable of distinguishing Covid-inspired papers from other publications. See the table of predicted topics and the words with the highest conditional probability for each topic in the following table. We define a paper as Covid-related if its probability of belonging to the "Covid-19" topic is higher than 50\%. The Covid index for each scholar is calculated as:

\begin{equation}
    \nonumber
    \text{Covid Index} = \frac{\text{Number of Covid-related Publications in 2019-2021}}{\text{Number of Publications in 2019-2021}}
\end{equation}

The mean Covid index for economic scholars is 0.11, meaning that on average, 11\% of the papers written by an economics faculty member in 2019-2021 are inspired by the pandemic.

\begin{table}[!htbp] \centering 
  \caption{Predicted Topics of Economic Publications in 2019-2021}
  \setlength\tabcolsep{2pt}
\begin{tabular}{llc}
\\[-1.8ex]\hline \hline \\[-1.8ex] 
\\[-1.8ex] \multicolumn{1}{c}{Topic} & \multicolumn{1}{c}{Key Words} & \multicolumn{1}{c}{Percentage(\%)} \\ 
\hline \\ [-1.8ex]
  Covid-19 & Covid-19, Health, Pandemic, Impact, Effects & 26.05 \\
  Macro/Finance/International & Financial, Trade, Monetary, Policy, Experimental & 19.97 \\ 
  Micro/Theory/Metrics/IO & Market, Theory, Labor, Learning, Estimation & 27.81 \\
  Development/Social Science/Others & Inequality, Gender, Mobility, Work, Review & 26.17 \\
 \hline \\[-1.8ex] 
\hline \\[-1.8ex]
\end{tabular} 
\end{table}

Social interaction matrix $\mathbf{G}$ is a row-normalized version of the adjacency matrix $\mathbf{W}=\left[w_{ij}\right]$, in which $w_{ij}=1$ if scholar $i$ and $j$ has collaborated on at least 2 papers within the 3-year period and $0$ otherwise. We rule out the case in which two scholars coauthored a single paper to focus on stable relationships. For example, if there are three scholars $i$, $j$ and $l$, scholar $i$ has worked with both $j$ and $l$ on more than one paper during the 3-year period. Then $\mathbf{G}$ is specified in the following way:

\begin{equation}
\nonumber
\mathbf{G} = \left[ 
  \begin{array}{ccc}
    0 & \frac{1}{2} & \frac{1}{2} \\
    1 & 0 & 0 \\
    1 & 0 & 0 \\ 
  \end{array}
\right]
\end{equation}

Each agent considers the average characteristics of his or her coauthors. For example, for the payoff function $U_i(y_i)$ in the above example, $\bar{y}_i=\frac{1}{2}y_j+\frac{1}{2}y_l$. Peer effect $\lambda$ indicates how much "peer pressure" a scholar may have due to the difference between the scholar's own productivity and the average productivity of the co-authors.

To account for the endogeneity of network formation, in the first stage, we estimate scholars' unobserved characteristics using a dyadic Logit model. In the network formation model, at the school level, we control for homophily by considering whether scholars are from the same department and the same US News Ranking category (Top 10, 11-20, 21-30, 31-40, 41-50). At the individual level, we consider differences between the two scholars in terms of academic experience in years, total citations up to the first year of the period, average number of publications each year during the previous three years, and total number of publications. We include dummies to indicate whether the dyad involves at least one female author and whether it involves at least one African American scholar. We also control for the two scholars' common research interests, measured by the number of fields listed on both scholars' websites. In the second stage, we include individual fixed effects as additional control variables to evaluate the peer effect.

In practice, the value of $\bar{R}$ needs to be specified by the researcher in advance. Following the suggestion of the author, We experiment with the value of $\bar{R}$ by increasing it from $2$ until the change of the estimated parameters is not significant, or it reaches $\max(y)-2$. The estimation is done using the R package provided by the author.

\section{Empirical Results}

The estimated peer effect is significant for pre-Covid period from 2018 to 2019, but not significant for Covid times from 2020 to 2021. This estimation implies that economic scholars exhibited conformity in the number of publications before schools switched to virtual mode. However, there is an indication that while working from home, economic scholars collaborated with a more diverse group of co-authors; in terms of productivity, there were more collaborations among prolific scholars than those who publish less frequently.

Our results illustrate that scholars’ recent productivity, represented by their average number of publications over the previous three years, is an important predictor of future productivity. In other words, scholars who were more prolific in the past three years tend to be more productive in the future. We also find that compared to scholars with fewer than 2,000 total citations, those with more citations are likely to publish more. Scholars’ years in academia significantly affect their productivity during Covid times, with emerging scholars producing relatively more than senior scholars. In addition, gender plays a significant role only during the pre-Covid period.

Moreover, our results show productivity is also related to the economics sub-fields that scholars work in. For example, econometricians collaborated and published relatively fewer papers in 2018-2019. Scholars working in health economics are more prolific, and this effect is stronger during Covid times. The significant and positive Covid index implies that during 2019-2021, the higher the proportion of a scholar’s papers related to Covid, the more publications he or she would have. After controlling for the Covid index, the coefficient for scholars working in the field of health economics drops from $0.33$ to $0.30$.

Furthermore, regarding the contextual effect, during Covid times, collaborating with macroeconomists and established scholars who have 100-2,000 citations helped increase the number of publications.

\begin{table}[!htbp] \centering 
  \caption{Peer Effect on Scholars' Number of Publications in Non-Covid and Covid Times} 
\begin{tabular}{lccc} 
\\[-1.8ex]\hline \hline \\[-1.8ex] 
\\[-1.8ex] & \multicolumn{1}{c}{Pre-Covid} & \multicolumn{1}{c}{Covid} & \multicolumn{1}{c}{Covid + Covid Index} \\ 
\\[-1.8ex] & (2018-2019) & (2020-2021) & (2020-2021) \\ 
\hline \\[-1.8ex]
  $\lambda$ & 0.10$^{***}$ & 0.01 & 0.02 \\ 
  & (0.03) & (0.03) & (0.03) \\
 \hline \\[-1.8ex] 
\hline \\[-1.8ex] 
\textit{Note:}  & \multicolumn{3}{r}{$^{*}$p$<$0.1; $^{**}$p$<$0.05; $^{***}$p$<$0.01} \\ 
\end{tabular}
\end{table} 

\begin{table}[!htbp] \centering 
  \caption{Own Effects on Scholars' Number of Publications in Non-Covid and Covid Times} 
\begin{tabular}{lccc} 
\\[-1.8ex]\hline \hline \\[-1.8ex] 
\\[-1.8ex] & \multicolumn{1}{c}{Pre-Covid} & \multicolumn{1}{c}{Covid} & \multicolumn{1}{c}{Covid + Covid Index} \\ 
\\[-1.8ex] & (2018-2019) & (2020-2021) & (2020-2021) \\ 
\hline \\[-1.8ex]
  2-4 Publications Per Year & 0.40$^{***}$ & 0.28$^{***}$ & 0.27$^{***}$ \\ 
  & (0.07) & (0.06) & (0.06) \\
  5-9 Publications Per Year & 1.22$^{***}$ & 1.17$^{***}$ & 1.17$^{***}$ \\ 
  & (0.09) & (0.09) & (0.09) \\
  10+ Publications Per Year & 2.38$^{***}$ & 2.46$^{***}$ & 2.46$^{***}$ \\ 
  & (0.16) & (0.16) & (0.16) \\
  100-499 Citations & $-$0.15$^{*}$ & $-$0.21$^{**}$ & $-$0.21$^{**}$ \\ 
  & (0.08) & (0.09) & (0.09) \\
  500-1,999 Citations & $-$0.02 & $-$0.11 & $-$0.11 \\ 
  & (0.07) & (0.07) & (0.07) \\
  2,000-4,999 Citations & 0.48$^{***}$ & 0.38$^{***}$ & 0.37$^{***}$ \\ 
  & (0.09) & (0.10) & (0.10) \\
  5,000-9,999 Citations & 0.71$^{***}$ & 0.59$^{***}$ & 0.58$^{***}$ \\ 
  & (0.13) & (0.13) & (0.13) \\
  10,000-19,999 Citations & 0.77$^{***}$ & 0.86$^{***}$ & 0.84$^{***}$ \\ 
  & (0.15) & (0.14) & (0.14) \\
  20,000+ Citations & 1.81$^{***}$ & 1.53$^{***}$ & 1.52$^{***}$ \\ 
  & (0.18) & (0.16) & (0.16) \\
  Experience 10-20 Years & $-$0.01 & $-$0.12 & $-$0.11 \\ 
  & (0.08) & (0.08) & (0.08) \\
  Experience 20-30 Years & $-$0.18$^{*}$ & $-$0.28$^{***}$ & $-$0.27$^{***}$ \\ 
  & (0.09) & (0.10) & (0.10) \\
  Experience 30-40 Years & $-$0.12 & $-$0.47$^{***}$ & $-$0.46$^{***}$ \\ 
  & (0.10) & (0.11) & (0.11) \\
  Experience 40-50 Years & $-$0.05 & $-$0.30$^{**}$ & $-$0.29$^{**}$ \\ 
  & (0.12) & (0.13) & (0.13) \\
  Experience 50-60 Years & 0.25$^{*}$ & $-$0.36$^{**}$ & $-$0.35$^{**}$ \\ 
  & (0.15) & (0.15) & (0.15)) \\
  Experience 60+ Years & 0.27 & $-$0.39$^{**}$ & $-$0.39$^{**}$  \\ 
  & (0.21) & (0.18) & (0.18) \\
  Covid Index &  &  & 0.27$^{*}$ \\ 
  &  &  & (0.16) \\
  African American & 0.33 & $-$0.07 & $-$0.06 \\ 
  & (0.29) & (0.29) & (0.29) \\
  Female & $-$0.13$^{*}$ & 0.04 & 0.04 \\ 
  & (0.07) & (0.07) & (0.07) \\
  Field: Theory & $-$0.10 & 0.09 & 0.10 \\ 
  & (0.07) & (0.07) & (0.07) \\
  Field: Macro & $-$0.04 & $-$0.001 & 0.01 \\ 
  & (0.07) & (0.07) & (0.07) \\
  Field: Labor & $-$0.09 & $-$0.02 & $-$0.04 \\ 
  & (0.07) & (0.07) & (0.07) \\
  Field: Metrics & $-$0.22$^{***}$ & 0.04 & 0.05 \\ 
  & (0.07) & (0.07) & (0.07) \\
  Field: Industrial Organization & $-$0.02 & $-$0.06 & $-$0.06 \\ 
  & (0.08) & (0.08) & (0.08) \\
  Field: Development & $-$0.04 & 0.16$^{*}$ & 0.16$^{*}$ \\ 
  & (0.08) & (0.08) & (0.08) \\
  Field: Health & 0.21$^{**}$ & 0.33$^{***}$ & 0.30$^{***}$ \\ 
  & (0.10) & (0.10) & (0.10) \\
  Field: Finance & $-$0.13 & $-$0.01 & $-$0.01 \\ 
  & (0.09) & (0.09) & (0.09) \\
 \hline \\[-1.8ex] 
\hline \\[-1.8ex] 
\textit{Note:}  & \multicolumn{3}{r}{$^{*}$p$<$0.1; $^{**}$p$<$0.05; $^{***}$p$<$0.01} \\ 
\end{tabular}
\end{table} 

\begin{table}[!htbp] \centering 
  \caption{Contextual Effects on Scholars' Number of Publications in Non-Covid and Covid Times} 
  \setlength\tabcolsep{1pt}
\begin{tabular}{lccc} 
\\[-1.8ex]\hline \hline \\[-1.8ex] 
\\[-1.8ex] & \multicolumn{1}{c}{Pre-Covid} & \multicolumn{1}{c}{Covid} & \multicolumn{1}{c}{Covid + Covid Index} \\ 
\\[-1.8ex] & (2018-2019) & (2020-2021) & (2020-2021) \\ 
\hline \\[-1.8ex]
  Proportion of Coauthors with 2-4 Publications Per Year & 0.05 & 0.13 & 0.11 \\ 
  & (0.23) & (0.17) & (0.17) \\
  Proportion of Coauthors with 5-9 Publications Per Year & $-$0.31 & 0.31 & 0.28 \\ 
  & (0.33) & (0.32) & (0.32) \\
  Proportion of Coauthors with 10+ Publications Per Year & $-$0.59 & $-$0.27 & $-$0.34 \\ 
  & (0.57) & (0.54) & (0.54) \\
  Proportion of Coauthors with 100-499 Citations & 0.11 & 0.44$^{*}$ & 0.46$^{*}$ \\ 
  & (0.26) & (0.27) & (0.27) \\
  Proportion of Coauthors with 500-1,999 Citations & $-$0.02 & 0.37$^{*}$ & 0.36 \\ 
  & (0.21) & (0.20) & (0.20) \\
  Proportion of Coauthors with 2,000-4,999 Citations & $-$0.07 & $-$0.09 & $-$0.07 \\ 
  & (0.35) & (0.30) & (0.30) \\
  Proportion of Coauthors with 5,000-9,999 Citations & $-$0.20 & 0.15 & 0.16 \\ 
  & (0.41) & (0.38) & (0.37) \\
  Proportion of Coauthors with 10,000-19,999 Citations & $-$0.49 & 0.26 & 0.28 \\ 
  & (0.37) & (0.41) & (0.41) \\
  Proportion of Coauthors with 20,000+ Citations & $-$0.52 & 0.38 & 0.38 \\ 
  & (0.58) & (0.53) & (0.53) \\
  Proportion of Coauthors with 10-20 Years Experience & 0.22 & 0.15 & 0.16 \\ 
  & (0.30) & (0.22) & (0.22) \\
  Proportion of Coauthors with 20-30 Years Experience & $-$0.39 & 0.05 & 0.05 \\ 
  & (0.33) & (0.24) & (0.24) \\
  Proportion of Coauthors with 30-40 Years Experience & $-$0.39 & $-$0.28 & $-$0.26 \\ 
  & (0.35) & (0.28) & (0.28) \\
  Proportion of Coauthors with 40-50 Years Experience & $-$0.02 & $-$0.69$^{**}$ & $-$0.69$^{**}$ \\ 
  & (0.37) & (0.32) & (0.32) \\
  Proportion of Coauthors with 50-60 Years Experience & $-$0.14 & $-$0.49 & $-$0.48 \\ 
  & (0.42) & (0.39) & (0.39) \\
  Proportion of Coauthors with 60+ Years Experience & $-$1.09$^{**}$ & 0.48 & 0.50 \\ 
  & (0.43) & (0.40) & (0.40) \\
  Covid Index (Coauthors) &  &  & $-$0.35 \\ 
  & & & (0.45) \\
  African American (Coauthors) & 0.12 & $-$0.68 & $-$0.68 \\ 
  & (0.52) & (0.78) & (0.77) \\
  Female (Coauthors) & 0.08 & 0.01 & 0.02 \\ 
  & (0.19) & (0.19) & (0.19) \\
  Field: Theory (Coauthors) & $-$0.06 & $-$0.10 & $-$0.12 \\ 
  & (0.21) & (0.17) & (0.17) \\
  Field: Macro (Coauthors) & $-$0.01 & 0.35$^{**}$ & 0.33$^{**}$ \\ 
  & (0.15) & (0.16) & (0.16) \\
  Field: Labor (Coauthors) & $-$0.45$^{**}$ & 0.15 & 0.16 \\ 
  & (0.21) & (0.16) & (0.16) \\
  Field: Metrics (Coauthors) & $-$0.06 & $-$0.07 & $-$0.07 \\ 
  & (0.17) & (0.17) & (0.17) \\
  Field: Industrial Organization (Coauthors) & $-$0.32 & 0.16 & 0.16 \\ 
  & (0.23) & (0.27) & (0.26) \\
  Field: Development (Coauthors) & $-$0.20 & $-$0.04 & $-$0.03 \\ 
  & (0.19) & (0.20) & (0.20) \\
  Field: Health (Coauthors) & 0.33 & 0.05 & 0.07 \\ 
  & (0.27) & (0.28) & (0.28) \\
  Field: Finance (Coauthors) & $-$0.03 & $-$0.10 & $-$0.10 \\ 
  & (0.21) & (0.21) & (0.21) \\
 \hline \\[-1.8ex] 
\hline \\[-1.8ex] 
\textit{Note:}  & \multicolumn{3}{r}{$^{*}$p$<$0.1; $^{**}$p$<$0.05; $^{***}$p$<$0.01} \\ 
\end{tabular}
\end{table} 

\section{Conclusion}

In this study, we examine the influence of peer effects on the productivity of economic scholars in both pre-pandemic and during pandemic periods. The findings reveal that scholars tended to collaborate with others of similar level of productivity during the pre-pandemic period, but this conformity was not observed during the pandemic. Productivity during the previous three years helps predict productivity in the near future. Citation count is positively correlated with scholars' productivity in both periods, while academic experience only affects productivity during the pandemic. Female scholars were reported to publish less during the pre-pandemic period, but this effect vanishes during the pandemic. This paper contributes to the growing body of literature on peer effects and academic productivity, offering insights that can help inform policies aimed at fostering research collaboration and enhancing productivity within the academic community.

While our research offers important insights, it is not without limitations. For example, in the present study, we assess each economics scholar's productivity based on her/his number of publications. This measure may be questionable because it does not account for factors like the quality of the research and the type of publications. More comprehensive measures could include citation counts, journal impact factors, types of publications, altmetrics, etc. Although Google Scholar has several advantages, such as broader coverage of different types of publications and up-to-date information, it may offer inaccurate author profiles and results of inconsistent accuracy \citep{falagas2008comparison}. Future research may benefit from incorporating additional data sources from journal websites or other online platforms to ensure a more comprehensive and accurate assessment of scholars' productivity and co-authorship patterns. Moreover, future research can use our findings to future explore the literature on Diversity, Equity, and Inclusion (DEI) in academia.

In addition, in our paper, we only investigate the peer effects within a four-year period. The Covid-19 pandemic has unprecedentedly affected the global academic community, potentially causing a paradigm shift as many scholars leave academia during and after the pandemic. Therefore, it is crucial to continue monitoring and studying the long-term implications of these collaboration patterns and their impact on research productivity. Further exploration of the factors that facilitate or hinder research collaboration during such crises can help guide the development of effective strategies and policies to support and strengthen the academic community in times of unprecedented challenges.

\newpage
\bibliography{references}

\end{document}